\documentclass[useAMS,usenatbib,usegraphicx]{mn2e}

\pagerange{\pageref{firstpage}--\pageref{lastpage}}
\pubyear{2010}

\newcommand{\be}{\begin{equation}}
\newcommand{\ee}{\end{equation}}

\begin{document}


\title[New constraints on Parity Symmetry from a re-analysis of the WMAP-7 low res power spectra]
{New constraints on Parity Symmetry from a re-analysis \\ of the WMAP-7 low resolution power spectra}
\author[A.Gruppuso et al.]
{A.~Gruppuso $^{1,2}$\thanks{E-mail: gruppuso@iasfbo.inaf.it }, F.~Finelli $^{1,2}$, P.~Natoli $^{3,4,5}$, F.~Paci $^{6}$, P.~Cabella $^{3}$
\newauthor A.~De Rosa $^{1}$, N.~Mandolesi $^{1}$
\\
$^1$ 
INAF-IASF Bologna, Istituto di Astrofisica Spaziale e Fisica Cosmica 
di Bologna \\
Istituto Nazionale di Astrofisica, via Gobetti 101, I-40129 Bologna, Italy \\
$^2$
INFN, Sezione di Bologna,
Via Irnerio 46, I-40126 Bologna, Italy 
\\
$^3$
Dipartimento di Fisica, Universit\`a di Roma Tor Vergata, Via della Ricerca Scientifica 1, 00133 Roma, Italy \\
$^4$
INFN, Sezione di Roma Tor Vergata, Via della Ricerca Scientifica 1, 00133 Roma, Italy \\
$^5$
ASI Science Data Center, c/o ESRIN, via G. Galilei, I-00044 Frascati, Italy \\
$^6$
Instituto de F\'isica de Cantabria (CSIC - Univ. de Cantabria), Avda. Los Castros s/n, 39005 Santander, Spain}

\label{firstpage}

\maketitle

\begin{abstract}
The Parity symmetry of the Cosmic Microwave Background (CMB) pattern as seen by WMAP 7 is tested
{\em jointly in temperature and polarization} at large angular scale. 
A Quadratic Maximum Likelihood (QML) estimator is applied to the WMAP 7 year
low resolution maps to compute all polarized CMB angular power spectra.
The analysis is supported by $10000$ realistic Monte-Carlo realizations.
We confirm the previously reported Parity anomaly for TT in the range $\delta \ell=[2,22]$ at $> 99.5\%$ C.L..
No anomalies have been detected in TT for a wider $\ell$ range (up to $\ell_{max}=40$).
No violations have been found for EE, TE and BB which we test here for the first time. The cross-spectra TB and EB are found to be consistent with zero. 
We also forecast {\sc Planck} capabilities in probing Parity violations on low resolution maps.
\end{abstract}



\section{Introduction}

The anisotropy pattern of the cosmic microwave background (CMB), 
measured by the Wilkinson Microwave Anisotropy Probe (WMAP), probes cosmology with unprecedented precision 
(see \cite{Larson:2010gs,Komatsu:2010fb} and references therein). 
WMAP data are largely consistent with the concordance $\Lambda$ cold dark matter ($\Lambda$CDM) model, but there are 
some interesting deviations from it, in particular on the largest angular scales \citep{Copi:2010na}.
See also \citep{Bennett:2010jb} for a critical point of view upon the subject.

A large number of papers dealing with these anomalies have been published in the last years. 
We list below those that are the most studied:
a) lack of power on large angular scales. The angular correlation function is found to be uncorrelated (i.e., consistent with zero) for angles larger than $60 ^{\circ}$. 
In \citep{Copi:2006tu,Copi:2008hw}, it was shown that this event happens in only $0.03\%$ of realizations of the concordance model.
b) Hemispherical asymmetries. It is found that the power in temperature coming separately from the two hemispheres (defined by the ecliptic plane) is unlikely asymmetric \citep{Eriksen:2003db,Hansen:2004vq}.
It has been confirmed in the WMAP 3 year and 5 year release \citep{Eriksen:2007pc,Hansen:2008ym,Hoftuft:2009rq} and it is present in the COBE data as well, although with lower significance. 
The temperature power spectra of the opposing hemispheres are inconsistent at $3\sigma$ to $4\sigma$ depending on the
range of multipoles considered. The asymmetry has been detected in low resolution maps \citep{Eriksen:2003db}, both in angular and multipoles space, but it extends to much smaller angular scales in the multipole 
range $\delta \ell = [2,600] $ \citep{Hansen:2008ym}.
At large angular scales the hemispherical asymmetry has been tested for the first time in polarization maps in  \citep{Paci:2010wp} making clear that this anomaly is evident only in intensity at WMAP sensitivity.
c) Unlikely alignments of low multipoles. An unlikely (for a statistically isotropic random field) alignment of the quadrupole and the octupole is found in \citep{Tegmark:2003ve,Copi:2003kt,Schwarz:2004gk,Weeks:2004cz,Land:2005ad}. 
Both quadrupole and octupole are shown to align with the CMB dipole \citep{Copi:2006tu}. Other unlikely alignments are described in \citep{Abramo:2006gw,Wiaux:2006zh,2007MNRAS.381..932V,Gruppuso:2010up}
and a test for detecting foreground residuals in the alignement estimators is present in \citep{Gruppuso:2009ee}.
d) Non-Gaussianity. \cite{Vielva:2003et} detected a localized non-Gaussian behavior in the southern hemisphere (called Cold Spot) using a wavelet analysis technique (see also \cite{Cruz:2004ce,Cruz:2009nd}).
Large scales non-Gaussian analyses can be found in \citep{Bernui:2009wq,Bernui:2010jt}.
See also \cite{Pietrobon:2009qg} where the needlet formalism has been applied to the WMAP 5 year data, looking for evidence of non-Gaussianity in the bispectrum of the needlet amplitudes.
e) Parity asymmetry. 
It has been suggested in \citep{Land:2005jq} that an estimator
built upon the point parity symmetry might be used as a practical tool
for detecting foregrounds.
In particular these authors consider whether the observed low 
CMB quadrupole in temperature could more generally signal odd point-parity, i.e. suppression of even multipoles.
However they claim that WMAP dataset never supports parity preference beyond the meagre $95 \%$ confidence level.
Later, \cite{Kim:2010gf} found that the Parity symmetry in the temperature map of WMAP 3 and 5 year data is anomalous at the level of 4 out of 1000 in the range $\delta \ell = [2,18]$. 
This analysis have been repeated in the WMAP 7 year data confirming the anomaly at same level for a slightly wider range $\delta \ell = [2,22]$ \citep{Kim:2010gd}.

In this paper we address the issue of the parity asymmetry estimating the angular power spectra (APS) of the WMAP 7 year data at large angular scales by an 
optimal and unbiased estimator such as the quadratic maximum likelihood (QML) estimator. 
The same estimator is used to analyze $10000$ simulated maps in which the noise is extracted from the low resolution noise covariance matrix
of the WMAP 7 year data. 
This approach is novel since the estimates of the $C_{\ell}$
are made {\it jointly} in temperature and polarization allowing for a global, more robust estimate of all six CMB spectra.
In particular, it allows to extend to polarization an analysis that has been performed only for temperature so far.

The used implementation of the QML is called {\it BolPol} and it has been already adopted 
in \citep{Gruppuso2009} to compute the APS of the low resolution WMAP 5 year data and in  \citep{Paci:2010wp}
where the hemispherical power asymmetries have been studied for the same data set.

The paper is organized as follows. In Section \ref{description} we describe
the performed analysis, introducing the property of symmetry we expect the CMB maps to have
in Subsection \ref{introparity}, providing the algebra of QML estimator in Subsection \ref{APS}, 
specifying the used data set and the performed simulations in Subsection \ref{dataset}
and defining the suitable estimators in Subsection \ref{estimators}.
Results are given in Section \ref{results} and forecasts for {\sc Planck}
are provided in Section \ref{forecast}. Conclusions are drawn in Section \ref{conclusions}.

\section{Description of the analysis}
\label{description}

\subsection{Introduction}
\label{introparity}

All-sky temperature maps, $T(\hat n)$, are usually expanded in terms of Spherical Harmonics 
$Y_{\ell m}(\hat n)$, with $\hat n$ being a direction in the sky,
namely depending on the couple of angles $(\theta, \phi)$:
\begin{equation}
a_{T, \ell m} = \int d\Omega \, Y^{\star}_{\ell m}(\hat n) \, T(\hat n) 
\label{almT}
\, ,
\end{equation}
where $a_{T, \ell m}$ are the coefficients of the Spherical Harmonics expansion
and $d \Omega = d \theta d \phi \sin \theta$. 
Under reflection (or Parity) symmetry ($\hat n \rightarrow -\hat n$), these coefficients behave as 
\begin{equation}
a_{T, \ell m} \rightarrow (-1)^{\ell} \,a_{T, \ell m}
\, .
\label{ParityT}
\end{equation}

Analogously for polarizations maps, taking into account the usual combination
of Stokes parameters ($Q(\hat n)$ and $U(\hat n)$)
\begin{eqnarray}
a_{\pm 2, \ell m} =  \int d\Omega \, Y^{\star}_{\pm 2, \ell m}(\hat n) \, ( Q(\hat n) \pm i U(\hat n)) \, ,
\label{QeU}
\end{eqnarray}
where $Y_{\pm 2, \ell m}(\hat n) $ are the Spherical Harmonics of spin $2$
and $a_{\pm 2, \ell m}$ are the corresponding coefficients,
it is possible to show that under Parity
\begin{eqnarray}
a_{E, \ell m} \rightarrow (-1)^{\ell} \,a_{E, \ell m} , \, \, \, \, \, \, 
\label{ParityE} \\
a_{B, \ell m} \rightarrow (-1)^{\ell+1} \,a_{B, \ell m}
\label{ParityB}
\, ,
\end{eqnarray}
where
\begin{eqnarray}
a_{E, \ell m} = -(a_{2, \ell m} +a_{-2, \ell m} )/2 \, , \, \\
a_{B, \ell m} = -(a_{2, \ell m} -a_{-2, \ell m} )/ 2 i
\, .
\label{ParityEeB}
\end{eqnarray}

Eqs.~(\ref{ParityT}), (\ref{ParityE}) and (\ref{ParityB}) show that the cross-correlations
$C_{\ell}^{TB} = C_{\ell}^{EB} = 0$.

Further details can be found for example in \citep{Zaldarriaga:1997yt}, \citep{Zaldarriaga:1996xe} and explicit algebra is present in Appendix \ref{parityproperties}.


\subsection{Angular Power Spectra Estimation}
\label{APS}

In order to evaluate the APS we adopt the QML estimator,
introduced in \citep{tegmark_tt}  and extended to polarization in \citep{tegmark_pol}. 
In this section we describe the essence of such a method. Further details
can be found in \citep{Gruppuso2009}.

Given a map in temperature and polarization ${\bf x=(T,Q,U)}$, the QML provides estimates
$\hat {C}_\ell^X$ - with $X$ being one of $TT, EE, TE, BB,
TB, EB$ - of the APS as: 
\be
\hat{C}_\ell^X = \sum_{\ell' \,, X'} (F^{-1})^{X \, X'}_{\ell\ell'} \left[ {\bf x}^t
{\bf E}^{\ell'}_{X'} {\bf x}-tr({\bf N}{\bf
E}^{\ell'}_{X'}) \right]
\, ,
\ee
where the $F_{X X'}^{\ell \ell '}$ is the Fisher matrix, defined as
\be
\label{eq:fisher}
F^{\ell\ell'}_{X X'}=\frac{1}{2}tr\Big[{\bf C}^{-1}\frac{\partial
{\bf C}}{\partial
  C_\ell^X}{\bf C}^{-1}\frac{\partial {\bf C}}{\partial
C_{\ell'}^{X'}}\Big] \,,
\ee
and the ${\bf E}^{\ell}_X$ matrix is given by
\be
\label{eq:Elle}
{\bf E}^\ell_X=\frac{1}{2}{\bf C}^{-1}\frac{\partial {\bf C}}{\partial
  C_\ell^X}{\bf C}^{-1} \, ,
\ee
with ${\bf C} ={\bf S}(C_{\ell}^X)+{\bf N}$ being the global covariance matrix (signal plus noise contribution). 

Although an initial assumption for a fiducial power spectrum $C_{\ell}^X$ is needed, the QML method provides unbiased estimates of the power spectrum contained 
in the map regardless of the initial guess,
\be
\langle\hat{C}_\ell^X\rangle= \bar C_\ell^{X} \,,
\label{unbiased}
\ee
where the average is taken over the ensemble of realizations (or, in a practical test, over Monte Carlo 
realizations ex-tracted from $\bar C_\ell^{X}$).
On the other hand, the covariance matrix associated to the estimates,
\be
\langle\Delta\hat{C}_\ell^X
\Delta\hat{C}_{\ell'}^{X'} \rangle= ( F^{-1})^{X \, X'}_{\ell\ell'} \,,
\label{minimum}
\ee
does depend on the initial assumption for $C_\ell^X$: 
the closer the guess to the true power spectrum is, the closer are the error bars to minimum variance.
According to the Cramer-Rao inequality, which sets a limit to the accuracy of an estimator, Eq. (\ref{minimum}) tells us that 
the QML has the smallest error bars.  The QML is then an `optimal' estimator.

\subsection{Data set and Simulations}
\label{dataset}

In this Section we describe the data set that we have considered. 
We use the temperature ILC map smoothed at $9.8$ degrees and reconstructed at HealPix\footnote{http://healpix.jpl.nasa.gov/}
 \citep{gorski} resolution $N_{side}=16$, 
the foreground cleaned low resolution maps and the noise covariance matrix in $(Q,U)$ publicly available at the LAMBDA website
\footnote{http://lambda.gsfc.nasa.gov/} for the frequency channels Ka, Q and V as considered by \cite{Larson:2010gs} for the low $\ell$ analysis.
These frequency channels have been co-added as follows \citep{Jarosik:2006ib}
\be
m_{tot} = C_{tot} (C_{Ka}^{-1} m_{Ka} + C_Q^{-1} m_Q+ C_V^{-1} m_V)
\, ,
\ee
where $m_{i}$, $C_i$ are the polarization maps  and covariances (for i=Ka, Q and V) and
\be
C_{tot}^{-1} = C_{Ka}^{-1} + C_{Q}^{-1} +C_{V}^{-1} 
\, .
\ee
This polarization data set has been extended to temperature considering the ILC map.
We have added to the temperature map a random noise realization with variance of $1 \mu K^2$ as suggested in \cite{dunkley_wmap5}.
Consistently, the noise covariance matrix for TT is taken to be diagonal with variance equal to $1 \mu K^2$.




We have also performed a Monte-Carlo simulations in order to assess the significance of our results.
A set of $10000$ CMB + noise sky realizations has been generated: the signal extracted from the WMAP 7 years best fit model, the noise through a Cholesky decomposition of the noise
covariance matrix. We have then computed the APS for each of the $10000$ simulations by means of {\it BolPol} and build two figures of merit as explained in the next subsection.

Two masks are considered: one for T and one for Q and U. Monopole and dipole have been subtracted from the observed ILC map through the HealPix routine {\sc remove-dipole} 
\citep{gorski}.

\subsection{Estimators}
\label{estimators}

We define the following quantities
\be
C^{X}_{+/-} \equiv {1\over {(\ell_{max}-1)}} \, \sum_{\ell=2,\ell_{max}}^{+/-} {\ell (\ell + 1) \over{2 \pi }} \, \hat{C}^{X}_{\ell} 
\label{C+-}
\ee
where $\hat{C}^{X}_{\ell}$ are the estimated APS obtained with {\it BolPol} for the power spectrum X = TT, TE, EE and BB.
The sum is meant only over the even or odd $\ell$ (and this is represented respectively by the symbol $+$ or $-$) with $\ell_{max} \ge 3$. 

Therefore, two estimators can be built from Eq.~(\ref{C+-}) as follows: 
the ratio $R^X$, as performed in \citep{Kim:2010gf} or \citep{Kim:2010gd},
\be
R^X = C^X_+/C^X_- \, ,
\label{ratioestimator}
\ee
and, in analogy to what performed for the hemispherical symmetry in \citep{Paci:2010wp}, the difference $D^X$
\be
D^X=C^X_+ - C^X_- \, ,
\label{diffestimator}
\ee
of the two aforementioned quantities. 
In the following, we drop the index $X$ for $R$ and $D$ specifying every time we use them which is the spectrum they refer to.
  
For our application to WMAP data, both estimators have been considered for the TT spectrum but only the second one for the other spectra (EE, TE and BB).
This is due the unfavorable signal-to-noise ratio of the WMAP data in polarization.

For X = TB and EB we simply use the average power
\be
C^{X} \equiv {1\over {(\ell_{max}-1)}} \, \sum_{\ell=2,\ell_{max}} {\ell (\ell + 1) \over{2 \pi }} \, \hat{C}^{X}_{\ell} 
\, .
\label{CTBeEB}
\ee

\section{Results}
\label{results}

APS for TT, TE and EE are given in Fig.~1
and for BB, TB and EB in Fig.~2.
In these Figures we display both the {\it BolPol} estimates of the WMAP 7 year maps (black symbols) and the Monte-Carlo realizations from $\ell=2$ to $32$ (blue symbols).
For TT and TE we display also the spectra provided by the WMAP team (red symbols) \footnote{Only these spectra are available on http://lambda.gsfc.nasa.gov/. 
Note that the WMAP TT estimates are obtained through a Maximum Likelihood code applied to the ILC map, whereas
the WMAP TE estimates are obtained through a pseudo-$C_{\ell}$ method where the foreground 
cleaned V band is considered for Temperature and foreground cleaned Q and V bands are taken into account for Polarization.}.

\begin{figure*}
\includegraphics[width=15cm,height=10cm,angle=0]{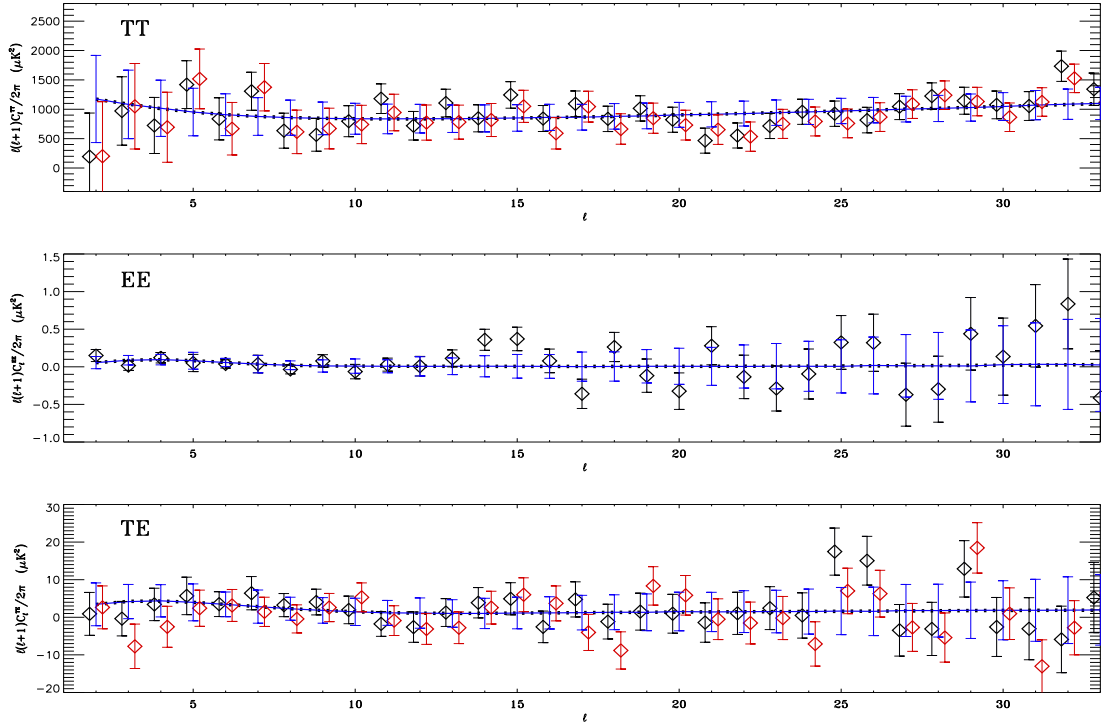}
\label{figurauno}
\caption{Estimates of TT (upper panel), EE (middle panel) and TE (lower panel) APS.
Dotted lines stand for the Fiducial Power Spectrum, taken to be the best fit of the WMAP 7 year data.
Blue lines with blue error bars represent the average and the standard deviation of a Monte Carlo made of $10000$ sky realizations
in which the full noise covariance is taken into account. Each realization has been analyzed with the {\it BolPol} code.
Red symbols are for the WMAP 7 year estimates as provided at the following web site http://lambda.gsfc.nasa.gov/.
Note that the WMAP TT estimates are obtained through a Maximum Likelihood code applied to the ILC map, whereas
the WMAP TE estimates are obtained through a pseudo-$C_{\ell}$ method where the foreground 
cleaned V band is considered for Temperature and foreground cleaned Q and V bands are taken into account for Polarization. 
Black symbols are for the {\it BolPol} WMAP 7 year estimates where the black error bars are given through the Fisher matrix.
Adopted resolution: $N_{side}=16$.}
\end{figure*}

\begin{figure*}
\includegraphics[width=15cm,height=10cm,angle=0]{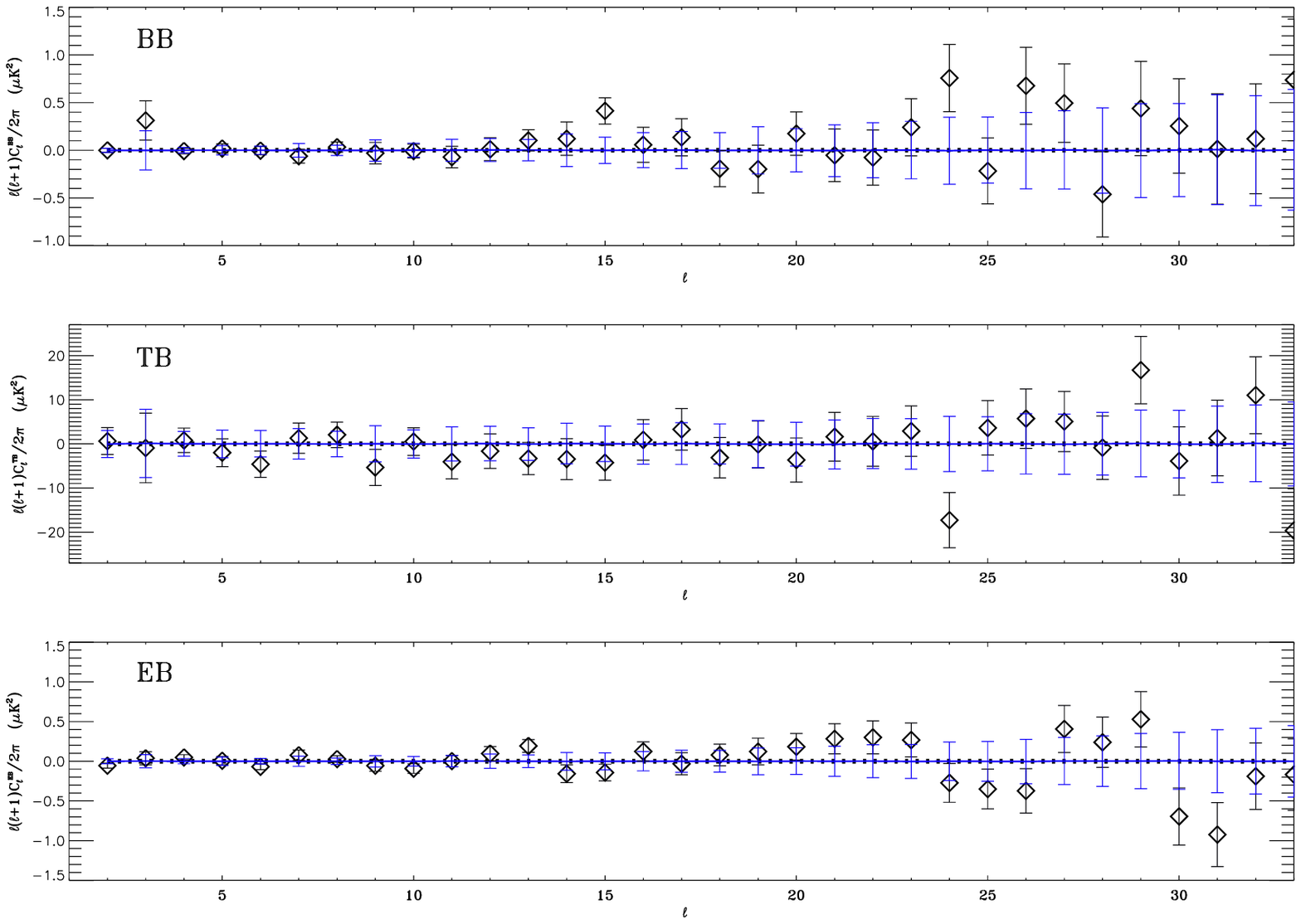}
\label{figuradue}
\caption{Estimates of BB (upper panel), TB (middle panel) and EB (lower panel) angular power spectrum.
Conventions as in Figure 1.}
\end{figure*}

In Fig. 3 we show the estimator $R$ and $D$ for TT averaged in $2 \leqslant \ell \leqslant 22$ and in $2 \leqslant \ell \leqslant 33$.
The probability to obtain a smaller value than the WMAP one is
$0.47\%$ for R in the range $\delta \ell = [2,22]$ and $3.17\%$ in the range $\delta \ell = [2,33]$.
For the D estimator the probability is $0.63\%$ in the range $\delta \ell = [2,22]$ and $3.17\%$ in the range $\delta \ell = [2,33]$.
The upper left panel of Fig.~3 recovers the same level of anomaly claimed in \cite{Kim:2010gd}.

\begin{figure}
\includegraphics[width=4.0cm,angle=0]{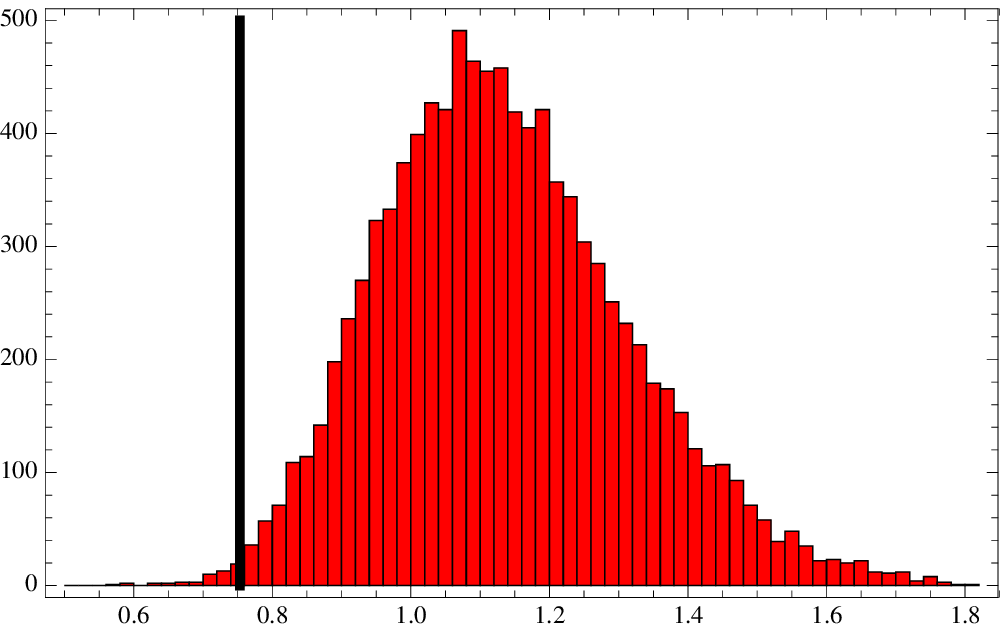}
\includegraphics[width=4.0cm,angle=0]{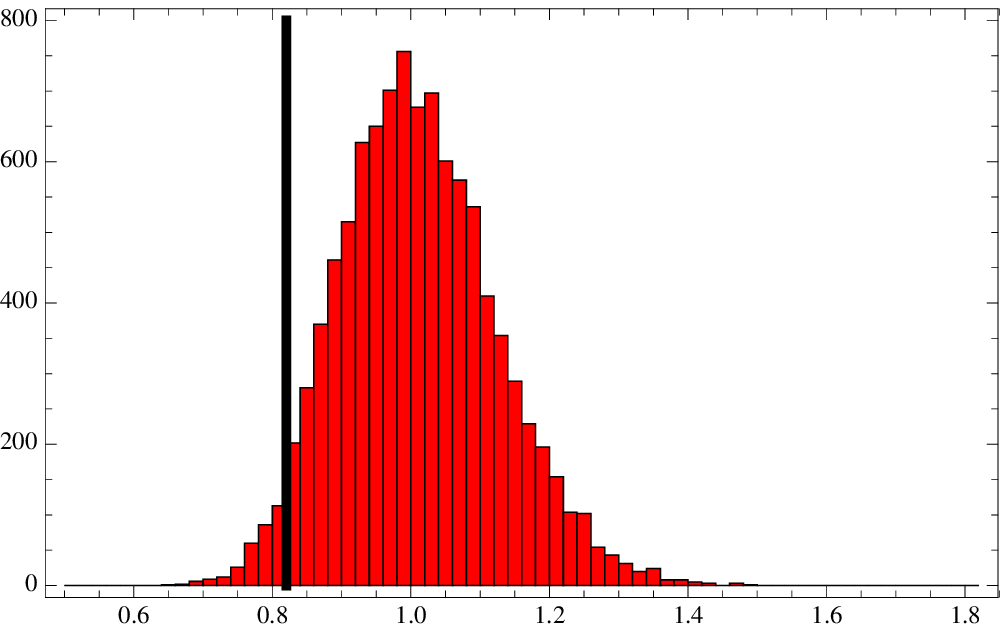}

\includegraphics[width=4.0cm,angle=0]{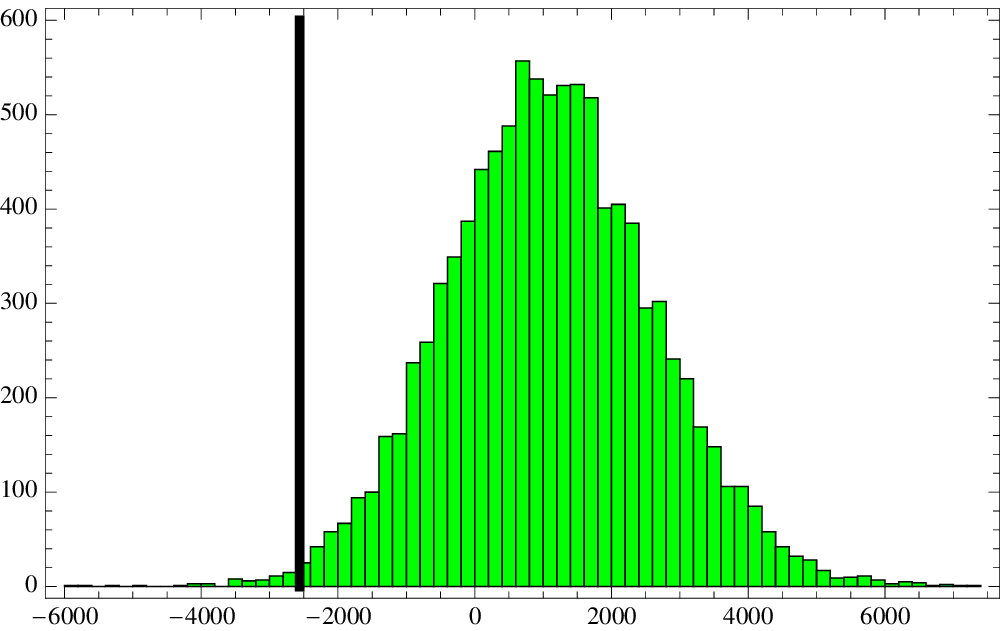}
\includegraphics[width=4.0cm,angle=0]{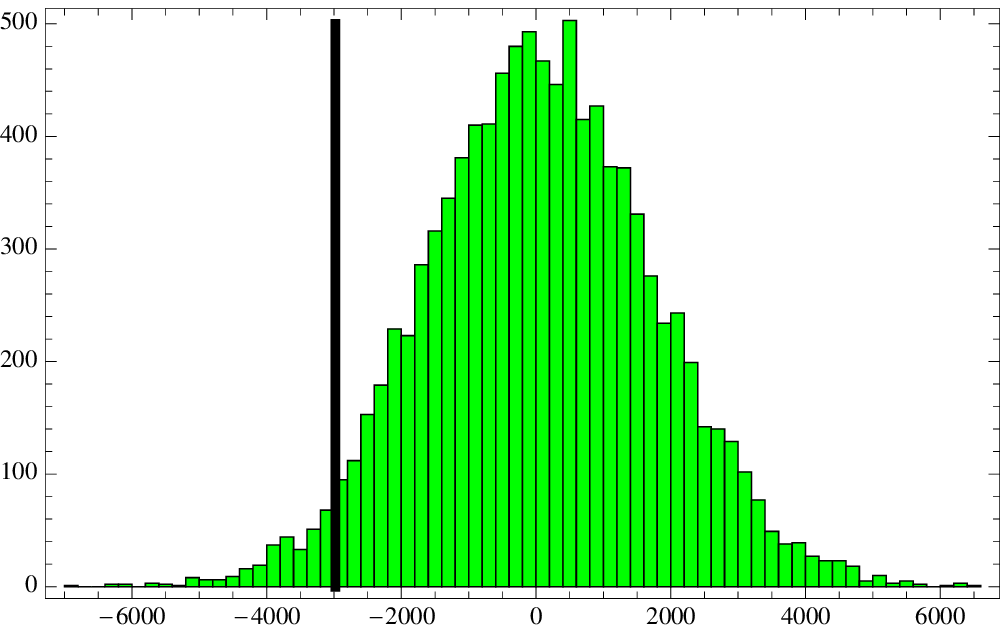}
%
\label{figura3}
\caption{TT. Counts (y-axis) vs the estimator (x-axis). Upper histograms: Ratio for the range $\delta \ell = [2,22]$ (left panel) and for the range $\delta \ell = [2,33]$ (right panel). 
Lower histograms: Difference for the range $\delta \ell = [2,22]$ (left panel) and for the range $\delta \ell = [2,33]$ (right panel).
Units for the estimator D are $\mu K^2$.
The vertical line stands for  the WMAP 7 year value.}
\end{figure}

In Fig.~4 we plot the percentage related to the WMAP 7 year Parity anomaly for TT versus $\ell_{max}$ in the range $10-40$ for the two considered estimators.
As evident there is not a single $\ell_{max}$ for which the TT anomaly shows up, but rather a characteristic scale,
see also \citet{Kim:2010gd}. For the estimator of Eq.~(\ref{ratioestimator}) the percentage anomaly is well below $1\%$ for almost any choices of $\ell_{max}$ in the range $[15,25]$
\footnote{Only for $\ell_{max} = 21$ the estimator of Eq.~(\ref{ratioestimator}) exhibits a percentage which is of the order of $1\%$.}. 
As also shown in Fig.~4 the estimator of Eq.~(\ref{diffestimator}) follows closely the other estimator although it is slightly less sensitive. 
Therefore, we find a whole multipole range, rather than a single $\ell_{max}$ value, where the WMAP 7 parity anomaly holds. 
This dims significantly the case for posterior biasing.
\begin{figure}
\includegraphics[width=8.0cm,angle=0]{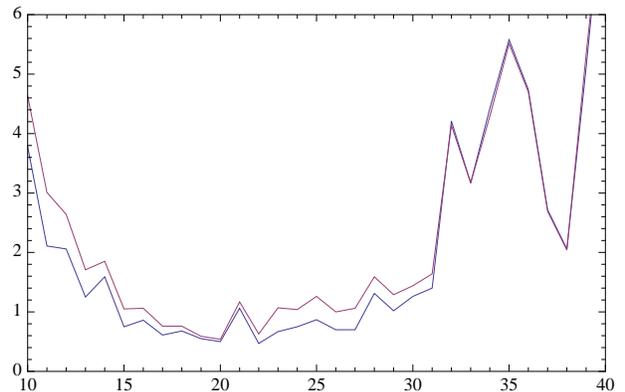}
\label{figura4}
\caption{TT. Percentage of the WMAP 7 year value (y-axis) vs $\ell_{max}$ (x-axis). 
Blue line is for the ratio and the red line for the difference.
This analysis shows that there is no single $\ell_{max}$ for which the TT anomaly shows up, but rather suggests the existence of a characteristic scale,
see also \citet{Kim:2010gd}.}
\end{figure}


In Table \ref{tableprobabilities2} we provide the results for EE, TE and BB. 
As mentioned above, only $D$  is considered and computed for the four following multipoles range
$\delta \ell = [2,4]$, $[2,8]$, $[2,16]$ and $[2,22]$.
No anomalies have been found and compatibility with Parity symmetry is obtained. 

In Table \ref{tableprobabilities3} we provide the results for EB and TB
where the estimator $C$ is considered and computed for the same aforementioned four multipoles range.
Both the spectra are well consistent with zero. Only the EB spectrum shows a mild anomaly in
the range $\delta \ell = [2,22]$ at the level of $97.7\%$. This is due to five estimates from $\ell=18$ to $\ell=22$ 
that are systematically larger than zero. When these points are excluded this
mild anomaly drops. For example in the range $\delta \ell = [2,16]$ the probability to obtain a smaller value
than the WMAP one is $55.35\%$. The latter two estimators are shown in Fig.~5.

\begin{table}
\caption{Probabilities (in percentage) to obtain a smaller value than the WMAP 7 year one} 
\centering 
\begin{tabular}{c c c c c} 
\hline\hline 
 D &  $\delta \ell$ = [2,4] & $\delta \ell$ = [2,8] &  $\delta \ell$ = [2,16] & $\delta \ell$ = [2,22] \\ [0.5ex] 
\hline 
EE & 93.09 & 76.21 & 44.27 & 46.61  \\ 
TE & 56.35 & 38.88 & 24.79 & 22.77 \\
BB & 7.97 & 13.42 & 11.70 & 44.31 \\ [1ex] 
\hline 
\end{tabular}
\label{tableprobabilities2} 
\end{table}

\begin{table}
\caption{Probabilities (in percentage) to obtain a smaller value than the WMAP 7 year one} 
\centering 
\begin{tabular}{c c c c c} 
\hline\hline 
 C &  $\delta \ell$ = [2,4] & $\delta \ell$ = [2,8] &  $\delta \ell$ = [2,16] & $\delta \ell$ = [2,22] \\ [0.5ex] 
\hline 
TB & 51.78 & 39.42 & 6.71 & 10.55  \\ 
EB & 62.73 & 69.83 & 55.35 & 97.70 \\ [1ex] 
\hline 
\end{tabular}
\label{tableprobabilities3} 
\end{table}

\begin{figure}
\includegraphics[width=4.0cm,angle=0]{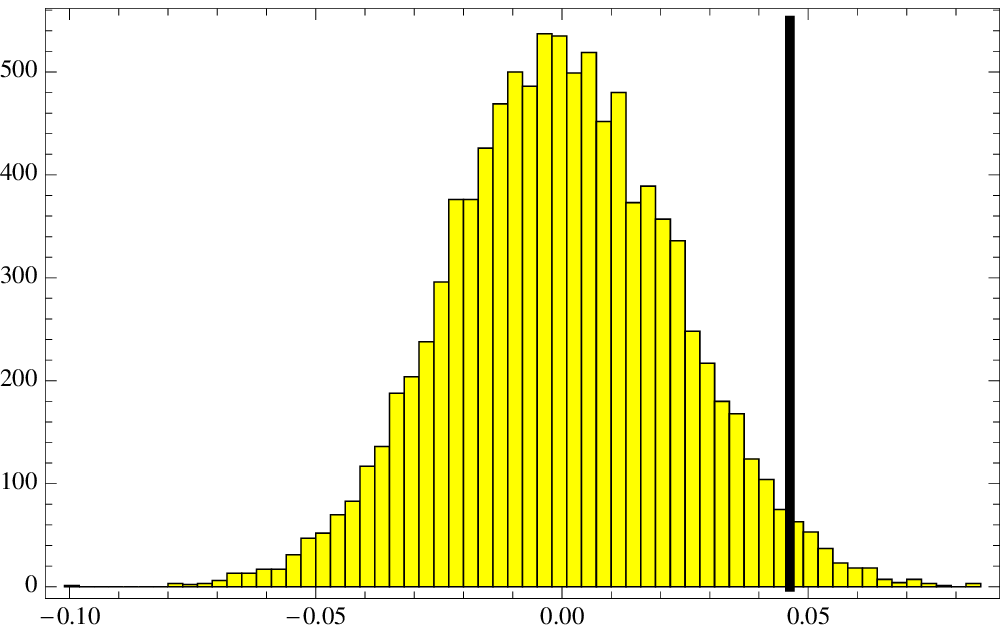}
\includegraphics[width=4.0cm,angle=0]{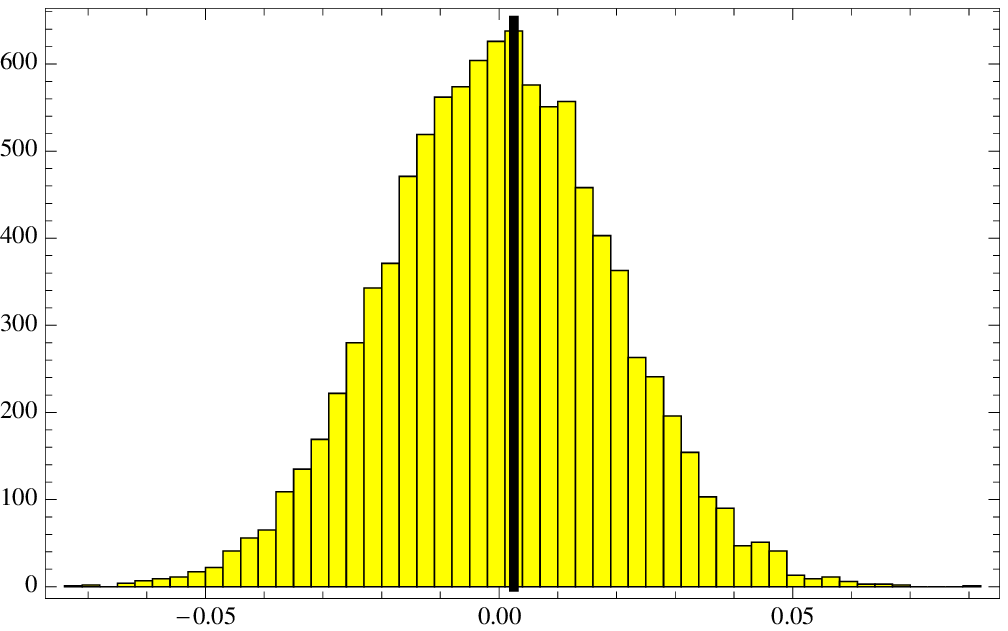}

%
\label{figura3}
\caption{EB. Counts (y-axis) vs the estimator C (x-axis). Distribution of $C$ for $\delta \ell = [2,22]$ (left panel) and $\delta \ell = [2,16]$ (right panel). 
Units are $\mu K^2$. The vertical line stands for  the WMAP 7 year data.}
\end{figure}

\subsection{Ka channel results}
We have considered those WMAP products that are already foreground reduced (see subsection \ref{dataset}).
However in order to test the case of a significant foreground contamination we have restricted our analysis to the Ka band since
such a channel is expected to be more polluted by synchrotron and free-free emission than the others. 
Despite the lower signal-to-noise ratio our analysis for the Ka channel is fully consistent with that of the entire dataset: no anomaly is detected in polarization, 
while in temperature we do confirm its existence at the same level.
These results, in our view, restrict (but of course not fully exclude) the chance for a significant foreground contamination.
The lower signal-to-noise ratio shows up in an increase of the standard deviations associated to the probability distribution functions of the estimators.
For example for the Ka band data set the standard deviation associated to the estimator D of EE in the $\ell$ range $[2,22]$
grows 3.7 times with respect to the one obtained with the entire data set.


\section{Planck Forecast}
\label{forecast}

In this Section we take into account the white noise level for $143$ GHz channel of the Planck mission \citep{bluebook} launched into space on the 14th of May of 2009.
As in \cite{Paci:2010wp}, we consider the nominal sensitivity of the Planck $143$ GHz channel, taken as representative of the results which can be obtained after the foreground cleaning from various frequency channels. 
The $143$ GHz channel has an angular resolution of $ 7.1^{\prime}$ (FWHM) and an average sensitivity of $6 \, \mu K \, \, (11.4 \, \mu K) $ per pixel - a square whose side is the FWHM size of the beam - in temperature (polarization), 
after $2$ full sky surveys. 
We assume uniform uncorrelated instrumental noise and we build the corresponding diagonal covariance matrix for temperature and polarization, from which, through Cholesky decomposition we are able to extract noise realizations.
For this low noise level we apply the same procedure adopted for the Monte-Carlo simulations in Subsection \ref{dataset}.
Thus, from the set of 10000 CMB + noise sky realizations, we find that:

\begin{itemize}
\item The T based estimators (both $R$ and $D$) do not change much since at large scale the APS for T is dominated by cosmic variance and not by the noise.
\item For EE, TE and BB it is possible to consider even the $R$ estimator. 
See for example Fig.~6 where the $R$ estimator is computed for EE in the range $\delta \ell = [2,22]$ (left panel) and $\delta \ell = [2,16]$ (right panel).
\item The standard deviations for the $D$ and $C$ are evaluated in Table \ref{tableprobabilities4} and \ref{tableprobabilities5} for $\delta \ell = [2,22]$
and compared to the WMAP 7 year ones.
\end{itemize}

\begin{table}
\caption{Standard deviation for the D estimator computed in the range $\delta \ell = [2,22]$. Units are $\mu K^2$.} 
\centering 
\begin{tabular}{c c c} 
\hline\hline 
 $\sigma_{D}$ &  WMAP 7 year & Planck \\ [0.5ex] 
\hline 
TT & 1517.17 & 1509.21   \\ 
TE & 20.19 & 9.08   \\ 
EE & 0.65 & 0.10   \\ 
BB & 0.69 & 0.04  \\ [1ex] 
\hline 
\end{tabular}
\label{tableprobabilities4} 
\end{table}

\begin{table}
\caption{Standard deviation for the C estimator computed in the range $\delta \ell = [2,22]$. Units are $\mu K^2$.} 
\centering 
\begin{tabular}{c c c} 
\hline\hline 
 $\sigma_{C}$ &  WMAP 7 year & Planck \\ [0.5ex] 
\hline 
TB & 0.95 & 0.19   \\ 
EB & 0.023 & 0.001  \\ [1ex] 
\hline 
\end{tabular}
\label{tableprobabilities5} 
\end{table}

\begin{figure}
\includegraphics[width=4.0cm,angle=0]{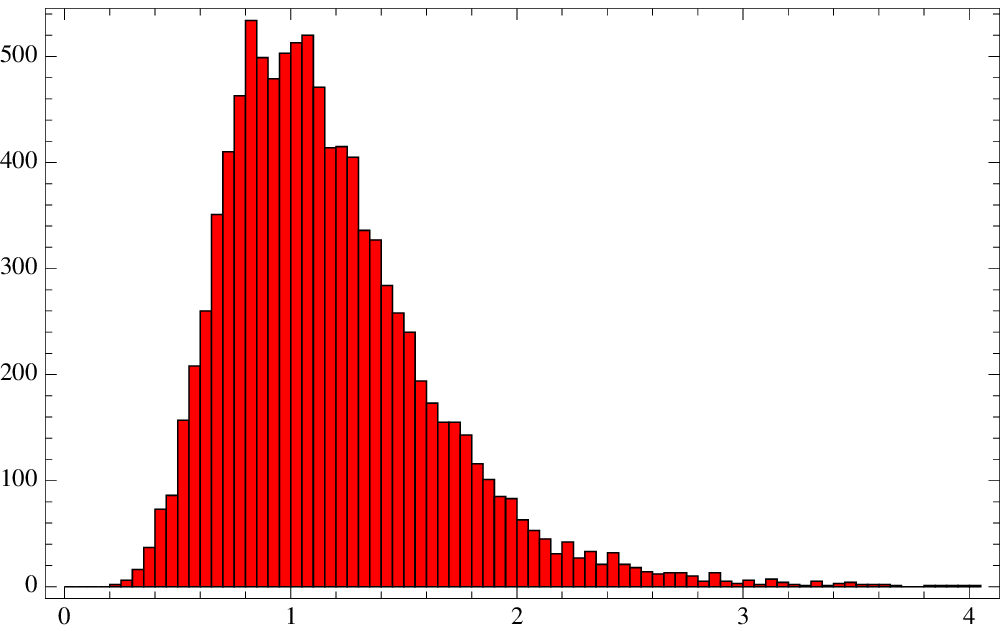}
\includegraphics[width=4.0cm,angle=0]{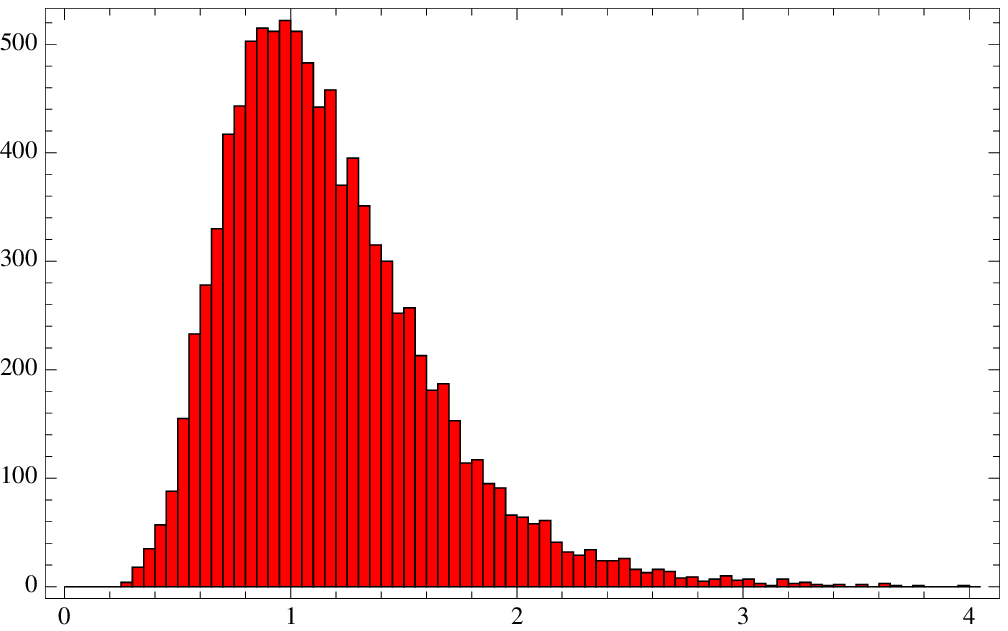}

%
\label{figura5}
\caption{EE. Counts (y-axis) vs the estimator R (x-axis). Distribution of $R$ for $\delta \ell = [2,22]$ (left panel) and $\delta \ell = [2,16]$ (right panel).}
\end{figure}



\section{Discussions and Conclusions}
\label{conclusions}
The Parity symmetry of the CMB pattern as seen by WMAP 7 year is tested {\em jointly in temperature and polarization} at large angular scale. 
We confirm the previously reported Parity anomaly for TT in the range $\delta \ell=[2,22]$ at $> 99.5\%$ C.L. \citep{Kim:2010gd}.
Our resolution 
allows to extend the angular range up to $\ell=40$ (see Fig.~4) finding a decrease of the anomaly for such a wider interval of multipoles.
No violations have been found for EE, TE and BB which are tested here for the first time with a method analogous to the one employed for TT. 
The cross-spectra TB and EB are found to be well consistent with zero.
The analysis has been performed through a Monte-Carlo made of $10000$ sky maps in which 
the CMB maps are extracted from the WMAP 7 year fiducial model and the noise maps are obtained
by Cholesky decomposition of WMAP 7 year noise covariance matrix for Ka, Q, V channels after foreground cleaning.
WMAP 7 year maps and each of Monte-Carlo simulations have been analyzed with {\it BolPol} that is our F90 implementation of the QML method.
As a byproduct the full APS of WMAP 7 year data is provided at large angular scale, see Figs.~1 and 2.

We have also forecasted {\sc Planck} capabilities in probing Parity violations. Considering $10000$ Monte-Carlo simulations we have evaluated the improvement with respect to WMAP 7 year sensitivity.
These are shown in Table 3 and 4.

It is still unknown whether this anomaly \footnote{Of course similar arguments apply to the other anomalies as well.} comes from fundamental physics or whether they are the residual 
of some unperfectly removed astrophysical foreground or systematic effect.
As shown in Section \ref{forecast}, {\sc Planck} data will allow one to build more precise estimators thanks to the high sensitivity of its instruments and the wide frequency coverage.
Moreover {\sc Planck} is observing the sky with a totally different scanning strategy with respect to WMAP, which is a benefit from the point of view of systematic effects analysis.
Thus {\sc Planck} data are awaited with great interest in order to confirm or discard these anomalies, making real the possibility to have more stringent constraints on
the $\Lambda$CDM model.

\section*{Acknowledgements}

We acknowledge the use of the BCX and SP6 at CINECA under the agreement INAF/CINECA and the use of computing facility at NERSC.
We acknowledge use of the HEALPix \citep{gorski} software and analysis package for
deriving the results in this paper.  
We acknowledge the use of the Legacy Archive for Microwave Background Data Analysis (LAMBDA). 
Support for LAMBDA is provided by the NASA Office of Space Science.
Work supported by ASI through ASI/INAF Agreement I/072/09/0 for
the Planck LFI Activity of Phase E2.

\appendix

\section{Parity Symmetry}
\label{parityproperties}

In this appendix we explicitly show the Eqs.~(\ref{ParityT}), (\ref{ParityE}) and (\ref{ParityB}).

\subsection{Temperature}

In analogy to the definition given in Eq.~(\ref{almT}) we write
\be
a^{(P)}_{T, \ell m} = \int d\Omega \, Y^{\star}_{\ell m}(\hat n) \, T^{(P)}(\hat n) 
\, ,
\label{almTParity}
\ee
where $T^{(P)}(\hat n) = P[T(\hat n) ]$ with $P$ being the parity operator. 
By definition $P[T(\hat n) ] = T(-\hat n)$, i.e. under reflection symmetry $\hat n \rightarrow -\hat n$ 
\footnote{This holds in 3 dimensions. Note that a Parity transformation in two dimensions is $(x,y) \rightarrow (x,-y)$ or $(x,y) \rightarrow (-x,y)$. 
In other words the transformation $(x,y) \rightarrow (-x,-y)$ is not a parity transformation since is equivalent to a rotation.} that in polar coordinates is equivalent to
\begin{eqnarray}
\theta &\rightarrow& \pi - \theta \, , 
\label{eqfortheta}
\\
\phi &\rightarrow& \phi + \pi 
\, .
\label{eqforphi}
\end{eqnarray}
Hence Eq.~(\ref{almTParity}) can be rewritten as
\be
a^{(P)}_{T, \ell m} = \int d\Omega \, Y^{\star}_{\ell m}(-\hat n) \, T(\hat n) 
\, ,
\label{almTParityprimo}
\ee
where the integration variables have been changed following Eqs.~(\ref{eqfortheta}),(\ref{eqforphi}) in order to absorb the minus in
the argument of $T$ \footnote{$\int d \Omega$ is invariant under Parity transformation. This is trivial to show in cartesian coordinates.}.
The Spherical Harmonics $Y_{\ell m}$ are related to Legendre functions $P^m_{\ell}$ as
\be
Y_{\ell m} (\theta, \phi) = \sqrt{\frac{2 \, \ell + 1}{4 \pi} \frac{(\ell -m)!}{(\ell + m)!}} P^m_{\ell}(\cos \theta) \, e^{i m \phi} 
\, ,
\ee
therefore
\be
Y_{\ell m} (\pi-\theta, \phi+\pi) = 
(-1)^m(-1)^{\ell +m}  Y_{\ell m} (\theta, \phi)
\, ,
\label{YforT}
\ee
where it has been used that 
\be 
P^m_{\ell}(\cos (\pi - \theta)) = (-1)^{(\ell +m)} P^m_{\ell}(\cos (\theta))
\, .
\ee
Replacing Eq.~(\ref{YforT}) in Eq.~(\ref{almTParityprimo}) gives
\be
a^{(P)}_{T, \ell m} = (-1)^{\ell}\int d\Omega \, Y^{\star}_{\ell m}(\hat n) \, T(\hat n) = (-1)^{\ell} \, a_{T, \ell m} 
\, ,
\label{almTParitysecondo}
\ee
that shows Eq.~(\ref{ParityT}).

\subsection{Polarization}

Let start considering the $E$ mode.
By definition
\be
a^{(P)}_{E,\ell m} = {-\frac{1}{2}} \left( a^{(P)}_{2,\ell m} + a^{(P)}_{-2,\ell m} \right)
\label{almEParity}
\, ,
\ee
where 
\be
a^{(P)}_{\pm 2, \ell m} = \int d\Omega \, Y^{\star}_{\pm 2,\ell m}(\hat n) \, [Q \pm iU]^{(P)}(\hat n) 
\, .
\label{qpiu}
\ee
Since $Q\pm iU$ is isomorphic to a bi-dimensional vector (see footnote 4) than
\begin{eqnarray}
[Q\pm iU]^{(P)}(\hat n) &=&  Q^{(P)}(\hat n) \mp  i U^{(P)}(\hat n)  \\
&=& Q(-\hat n) \mp  i U(-\hat n) 
\label{parityQeU}
\, .
\end{eqnarray}
Replacing Eqs.~(\ref{parityQeU}) and (\ref{qpiu})  in Eq.~(\ref{almEParity}) we find
\be
a^{(P)}_{E,\ell m} = - \! \!  \int \! \! d\Omega \left[ X^{\star}_{1,\ell m}(-\hat n) \, Q(\hat n)  -  X^{\star}_{2,\ell m}(-\hat n) iU(\hat n) \right]
\label{almEParityprimo}
\, 
\ee
where we have changed the integration variables considering Eqs.~(\ref{eqfortheta}),(\ref{eqforphi}) and where 
we have used the conventions of \citep{Zaldarriaga:1997yt} defining $X_{1/2, \ell m}$ as
\begin{eqnarray}
X_{1,\ell m} (\theta,\phi) = {-\frac{1}{2}} \left( Y_{2,\ell m} (\theta,\phi) + Y_{-2,\ell m} (\theta,\phi) \right) \, , 
\label{defX1}
\\
X_{2,\ell m} (\theta,\phi) = {-\frac{1}{2}} \left( Y_{2,\ell m} (\theta,\phi) - Y_{-2,\ell m} (\theta,\phi) \right)
\, .
\label{defX2}
\end{eqnarray}
Since in \citep{Zaldarriaga:1997yt} $X_{1/2, \ell m}$ are expressed in terms of Legendre polyniomials
it is easy to show that
\be
X_{1,\ell m}(\pi - \theta, \phi + \pi)  = (-1)^{\ell} X_{1,\ell m}(\theta, \phi)
\, ,
\label{X1sym}
\ee
and
\be
X_{2,\ell m}(\pi - \theta, \phi + \pi)  = (-1)^{\ell+1} X_{2,\ell m}(\theta, \phi)
\, .
\label{X2sym}
\ee
Replacing Eqs.~(\ref{X1sym}),(\ref{X2sym}) back into Eq.~(\ref{almEParityprimo})
and using again Eqs.~(\ref{defX1}),(\ref{defX2}) we find Eq.~(\ref{ParityE})
\be
a^{(P)}_{E,\ell m} = (-1)^{\ell} a_{E,\ell m}
\, .
\ee

Repeating the same steps but starting from
\be
a^{(P)}_{B,\ell m} = {-\frac{1}{2i}} \left( a^{(P)}_{2,\ell m} - a^{(P)}_{-2,\ell m} \right)
\label{almBParity}
\, ,
\ee
one finds Eq.~(\ref{ParityB}), i.e.
\be
a^{(P)}_{B,\ell m} = (-1)^{\ell + 1} a_{B,\ell m}
\, .
\ee






\end{document}